\documentclass[aps,pra,showpacs,twocolumn,amsmath,amssymb,groupedaddress]{revtex4-1}

\usepackage{graphicx}
\begin{document}


\title{Solution for (1+1) dimensional surface solitons in thermal nonlinear media}

\author{Xuekai Ma, Zhenjun Yang, Daquan Lu, Qi Guo, Wei Hu}

\email[Corresponding author's email address: ]{huwei@scnu.edu.cn}

\affiliation{Laboratory of Photonic Information Technology, South
China Normal University, Guangzhou 510631, P. R. China}


\date{\today}

\begin{abstract}
Analytical solutions for (1+1)D surface fundamental solitons in
thermal nonlinear media are obtained. The stationary position and the
critical power of surface solitons are obtained using this
analytical solutions. The analytical solutions are verified by
numerical simulations. The solutions for surface breathers and
their breathing period, and solutions for surface dipole and tripole solitons
are also given.
\end{abstract}


\maketitle


\section{introduction}
In nonlinear optics, an optical beam forms a soliton when the
self-focusing caused by the nonlinearity of the medium balances the
diffraction. Recently, people pay more attention to nonlocal
nonlinear media
\cite{Snyder-1997-Science,Krolikowski-2000-PRE,Conti-2004-PRL,Rotschild-2005-PRL}.
There are some types of solitons in nonlocal media, such as
multipole-mode solitons
\cite{Deng-2007-JOSAB,Xu-2005-OL,Rotschild-2006-OL},
Laguerre and Hermite solitons \cite{Buccoliero-2007-PRL}, and
vortex solitons \cite{Rotschild-2005-PRL,Alexander-2005-PRE,Briedis-oe-2005}.

Surface waves have been
studied generally in physics, chemistry and biology. They have been
used to study the surface properties of media and the
interaction between  media and optical beams. In the presence of nonlinearity, some kinds of surface solitons
have been found theoretically and experimentally, such as in local Kerr media \cite{Tomlinson-ol-1980,Mihalachea-pio-1989},
waveguide arrays \cite{Makris-ol-2005,Suntsov-prl-2006,Katashov-prl-2006},  photorefractive media \cite{Quirino-pra-1995,Cronin-Golomb-ol-1995,Aleshkevich-pre-2001}, and metamaterials \cite{Lazarides-pre-2008}

Nowadays,  surface solitons in nonlocal nonlinear media have been found
theoretically and experimentally \cite{Alfassi-2007-PRL}. An
interesting property of surface solitons is that a beam launched
away from the stationary position moves to the interface and
oscillates at its vicinity, even if the launch position is far away
from the stationary position. Now, varies types of surface
solitons have been found and studied, such as incoherent surface
solitons \cite{Alfassi-2009-PRA}, surface dipoles
\cite{Ye-2008-PRA,V.Kartashov-2009-OL} and surface vortices
\cite{Ye-2008-PRA}. Surface dipoles are found to be stable in the
entire existence domain. Surface vortices are found to exhibit
strongly asymmetric intensity and phase distributions. Bound states
of surface vortex solitons belong to a class of surface solitons
having no counterparts in bulk media \cite{Ye-2008-PRA}.

However, there are no analytical solutions for the family of
surface solitons due to the boundary conditions. We notice that the
optical field of surface solitons almost resides in the
nonlinear medium if the refractive index difference between two media
is large enough. It is reasonable to assume that the optical
intensity equals zero  at the interface with large refractive index
difference. In this paper, we obtain analytical solutions for the
family of surface solitons and surface breathers under this
assumption. Using the analytical solutions, we also obtain the
stationary position and the critical power of surface solitons,
and the breathing period of surface breathers. These results are
proved by numerical simulations.

\section{surface fundamental solitons }

\begin{figure}[b]
  \includegraphics[width=7.6cm]{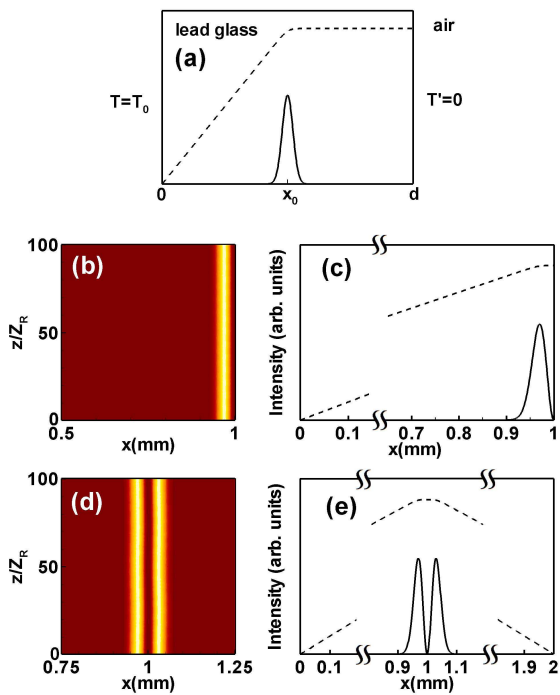}\\
   \caption{(Color online) (a) Sketch of a (1+1)D surface soliton. (b) Propagation
of a surface soliton, and (c) its intensity and nonlinear refractive
index profiles. (d) Propagation of a first-order Hermite-Gaussian
soliton in a bulk medium, and (e) its intensity and nonlinear
refractive index profiles. The beam width is $30$ $\mu$m
and $Z_R=k_1w_0^2$ is the Rayleigh distance. The widths of the media are
1 mm for the surface soliton, and 2 mm for the bulk soliton,
respectively. The solid line is the optical intensity profile and
the dashed line is the nonlinear refractive index profile.}
\label{fig1}
\end{figure}

First, we consider a (1+1)D sample sketched in Fig. \ref{fig1}(a)
following the experiment in Ref. \cite{Alfassi-2007-PRL}. The sample
width $d$ is $1$ mm. The interface ($x=d$) between a nonlocal
nonlinear medium (lead glass) with a refractive index $n_1=1.8$ and
a linear medium (air) with a refractive index $n_2=1.0$ is thermally
insulating, thus the temperature distribution $T$ satisfies the
boundary condition $\partial T/\partial x|_{x=d}=0$. The other side
of the sample is thermally conductive at the temperature $T_0$,
$T_0$ is the temperature in the absence of light, thus the boundary
condition is $T|_{x=0}=T_0$. The nonlinear refractive index change
($\Delta n$) of the sample satisfies the relation $\Delta
n=\beta\Delta T=\beta(T-T_0)$, where $\beta$ is the thermal
coefficient of the refractive index. $\Delta n$ satisfies the
conditions
\begin{equation}\label{200}
   \left.\frac{\partial (\Delta n)}{\partial x}\right|_{x=d}=0,  \,\,\,\,\,\,
   \Delta n|_{x=0}=0.
\end{equation}
$T$ satisfies the heat diffusion equation \cite{Rotschild-2005-PRL}
\begin{equation}\label{2}
    \kappa\frac{\partial^2T}{\partial x^2}=-\alpha I,
\end{equation}
where $\kappa$ is the thermal conductivity coefficient, $I$ is the optical
intensity, $\alpha$ is the absorption coefficient.

We consider the propagation of a TE polarized laser beam along $z$ axis in
the vicinity of the interface. In the nonlocal nonlinear medium
($0\leq x\leq d$), the slowly varying light field amplitude $U(x,z)$
(i.e. $I=|U|^2$) satisfies the nonlocal nonlinear Schr\"{o}dinger
equation,
\begin{equation}\label{1}
2ik_1\frac{\partial U}{\partial z}+\frac{\partial^2U}{\partial x^2}
+2k_1^2\left(\frac{\Delta n}{n_1}\right)U=0,
\end{equation}
where $k_1=n_1k_0 $, $k_0$ is wave number in vacuum and $|\Delta
n|\ll n_1$. In the linear medium ($x\geq d$), $U(x,z)$ satisfies
\begin{equation}\label{01}
2ik_1\frac{\partial U}{\partial z}+\frac{\partial^2U}{\partial x^2}
+(k_2^2-k_1^2)U=0,
\end{equation}
where $k_2=n_2k_0 $. Here we ignore the absorption of the medium, which is necessary for producing temperature changes and thermal nonlinearity. If the absorption is considered in propagation, the rigorous unchanged soliton does not exist in the nonlocal nonlinear medium due to the loss of the beam energy and the decreasing of the nonlinearity \cite{Huang-oc-2006,Cao-oc-2008}.  Fortunately the propagation distances in most experiments are short enough so that the influence of the absorption can be ignorable \cite{Rotschild-2005-PRL,Alfassi-2007-PRL,Alfassi-2009-PRA}. In order to show the stability of solitons in this paper, the propagation distances in numerical simulations are chosen to be 100-500 diffraction lengths, which are much longer than that in real experiments.

In Fig. \ref{fig1}(a), the solid line is the optical intensity
profile for the beam launched at $x=x_0$ and the dashed line is the
nonlinear refractive index which is not a bell shape. Obviously, the
optical beam will approach to the interface where the nonlinear
refractive index is larger, which presents the nonlinear attraction
of the interface. The optical beam will be reflected when it arrives
at the interface because the linear refractive index of the sample
is larger than that of the air. When the nonlinear attraction
balances the interface reflection, and simultaneously the nonlinear self-focusing balances
the diffraction, the beam propagates
stably along $z$ axis, which forms a surface soliton as shown in
Figs. \ref{fig1}(b) and \ref{fig1}(c).
Because the refractive index difference between two media is
 large enough, i.e. $n_1-n_2=0.8$, there is very little optical intensity
residing in the air. In the simulation, the ratio of the optical
intensity at the interface ($x=d$) to the maximum optical intensity
is approximately $10^{-4}$, so the optical intensity at the
interface can be negligible. Therefore we introduce the boundary conditions
\begin{equation}\label{003}
I|_{x=0}=0,  \,\,\,\,\,\,
I|_{x=d}=0,
\end{equation}
where $I|_{x=0}=0$ is valid because the optical beam is narrow and
far away from the left boundary. Using above conditions and Eq.
(\ref{200}) for the temperature distribution, one can completely
solve Eqs. (\ref{2}) and (\ref{1}) for the propagation of optical
beams in the nonlinear medium, i.e. $0\leq x \leq d$. Here the
physical influence of the linear medium, governed by Eq. (\ref{01})
with continuous conditions at the interface, is simplified to the
boundary condition $I|_{x=d}=0$ for the circumstance when the refractive
index difference between two media is large
enough.

To obtain analytical solutions for Eqs. (\ref{2}) and (\ref{1}),
we consider a (1+1)D first-order Hermite-Gaussian (HG) soliton in a
same bulk medium with the sample width $2d$, as shown in Figs. \ref{fig1}(d) and
\ref{fig1}(e). The optical intensity satisfies the boundary
conditions
\begin{equation}\label{0001}
I|_{x=0}=0, \,\,\,\,\,\,
I|_{x=2d}=0,
\end{equation}
because the optical beam is narrow and far away from the two boundaries. It also satisfies the condition
\begin{equation}\label{0003}
I|_{x=d}=0,
\end{equation}
because the first-order HG function is antisymmetric and launched at the sample
center ($x=d$).  The temperature $T$
satisfies the conditions $T|_{x=0}=T_0$ and $T|_{x=2d}=T_0$, because the two boundaries of the sample are thermally conductive at the temperature $T_0$.
Therefore the refractive index change satisfies the boundary
conditions
\begin{equation}\label{0004}
\Delta n|_{x=0}=0, \,\,\,\,\,\,
\Delta n|_{x=2d}=0.
\end{equation}
The refractive index change is symmetrical for a
symmetrical optical intensity, so it satisfies
\begin{equation}\label{0006}
\left.\frac{\partial(\Delta n)}{\partial x}\right|_{x=d}=0.
\end{equation}

The evolution of the first-order HG soliton in the bulk medium ($0\leq
x\leq 2d$) is governed by Eqs. (\ref{2}) and (\ref{1}) under the
boundary conditions Eqs. (\ref{0001}) and (\ref{0004}). Comparing
Eqs. (\ref{0003}) and (\ref{0006}) for the antisymmetric HG bulk soliton
with Eqs. (\ref{200}) and (\ref{003}) for the surface soliton, one can
find that the surface soliton is identical with the half part
($0\leq x\leq d$) of the first-order HG bulk soliton. Therefore, the
solution of the surface soliton can be obtained from
the solution of the first-order HG soliton in the bulk medium, namely,
\begin{equation}\label{3}
U(x)=U_0H_1\left(\frac{x-d}{w_0}\right)\exp\left[-\frac{(x-d)^2}{2w^2_0}\right]\exp(-ibz),
\end{equation}
where $H_1$ is the first-order Hermite polynomial, $w_0$ is the beam
width, $b$ is the propagation constant, $U_0$ is a constant, and $0\leq x\leq d$.

\begin{figure}[htbp]
  \includegraphics[width=7.6cm]{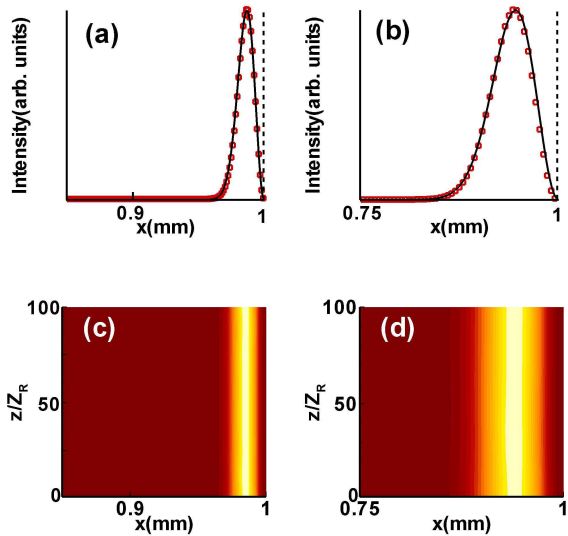}\\
  \caption{(Color online) The intensity profile of surface solitons for (a)
$w_0=15$ $\mu$m and (b) $w_0=60$ $\mu$m. The square symbols
represent the numerical solutions and the solid lines represent the
analytical solutions.  (c) and (d), propagations of surface solitons
using the analytical solution as incident profile corresponding to
(a) and (b), respectively. The sample width is 1 mm. The dashed
lines indicate the interface.} \label{ff2}
\end{figure}

Regarding the analytical solution as a trial solution, we will
find the numerical soliton profile using the iterative method based
on Eqs. (\ref{200})-(\ref{01}). In our simulations, the parameters
are $\alpha=0.01$ cm$^{-1}$, $T_0=25$ $^\circ$C,
$\kappa=6.37\times10^{-3}$ WK$^{-1}$cm$^{-1}$,
$\beta=1.4\times10^{-5}$ K$^{-1}$, $\lambda=514$ nm, which are the
same as those in Ref. \cite{Alfassi-2007-PRL}. The numerical
profiles are calculated without the assumption $I|_{x=d}=0$,
whereas all results of intensities at $x=d$ are small enough to be
approximated to zero. Figs. \ref{ff2}(a) and \ref{ff2}(b) show the
comparison of the optical intensity of the analytical solutions
(solid lines) with the numerical solutions (square symbols) for
$w_0=15$ $\mu$m [Fig. \ref{ff2}(a)] and $w_0=60$ $\mu$m [Fig.
\ref{ff2}(b)]. One can see that the analytical solutions are in good
agreement with the numerical solutions. Figs. \ref{ff2}(c) and
\ref{ff2}(d) show the propagations of surface solitons using the
analytical solutions Eq. (\ref{3}) as incident profiles. The
propagations are stable for a relative long distance, which verifies
our analytical solutions again.

The critical power of surface solitons can be obtained from the
analysis of bulk solitons. In the bulk medium, the nonlinear
refractive index in Eq. (\ref{1}) can be expanded at $x=d$, as
\cite{Snyder-1997-Science}
\begin{equation}\label{444}
\Delta n=(\Delta n)^0-\frac{(x-d)^2}{2}\delta^2(P_0),
\end{equation}
where $P_0$ is the input power of bulk solitons. The parameter
$\delta^2$ can be found to be proportional to the beam intensity at
$x=d$, or be proportional to $P_0/[\sqrt{\pi}w(z)]$, i.e.
\begin{equation}\label{445}
\delta^2(P_0)=\frac{\alpha\beta}{\kappa}\frac{P_0}{\sqrt{\pi}w(z)}.
\end{equation}
Then, Eq. (\ref{1}) reduces to a linear equation
\begin{equation}\label{446}
2ik_1\frac{\partial U}{\partial z}+\frac{\partial^2U}{\partial x^2}
-k_1^2\left(\frac{\alpha\beta
}{n_1\kappa}\right)\frac{P_0}{\sqrt{\pi}w(z)}(x-d)^2U=0.
\end{equation}
By substituting the solution of bulk solitons [same as Eq.
(\ref{3}) but $0\leq x\leq 2d$ ] into Eq. (\ref{446}), and from the coefficient of
each order term of $x$, we obtain the relation between the soliton
power and the beam width in the bulk medium,
\begin{equation}\label{5}
P_c=\frac{2\sqrt{\pi}n_1\kappa}{\alpha\beta k_1^2w_0^3}.
\end{equation}
The critical power of surface solitons is half of that of bulk
solitons, i.e. $P_s=P_c/2$.  The amplitude constant
$U_0=\sqrt{P_c}/(\sqrt{\pi}w_0)^{1/2}$, and the propagation constant
\begin{equation}\label{447}
b=\frac{3}{2k_1w_0^2}.
\end{equation}
The stationary position $x_s$, namely the maximum optical intensity
position of surface solitons, can  be also obtained,
\begin{equation}\label{4}
x_s=d-w_0.
\end{equation}
Now the analytical solution for fundamental surface solitons is
obtained completely. For example, in Fig. \ref{ff2}(c), the beam
width $w_0$ is $15$ $\mu$m, the critical power $P_s$ is $888.4$ W/cm
[in the simulation $P_s=892$ W/cm] and the stationary position is
$15$ $\mu$m away from the interface. In Fig. \ref{ff2}(d), the beam
width $w_0$ is $60$ $\mu$m, the critical power $P_s$ is $13.9$ W/cm
[in the simulation $P_s=14.2$ W/cm] and the stationary position is
$60$ $\mu$m away from the interface.

\begin{figure}[htbp]
  \includegraphics[width=7.6cm]{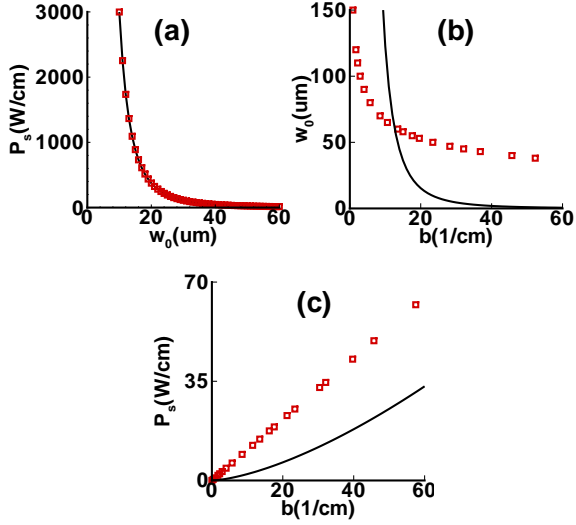}\\
  \caption{(Color online)
  (a) Relation between the critical power $P_s$ and the beam width $w_0$.  (b) Relation
between the beam width $w_0$ and the propagation constant $b$. (c) Relation
between the soliton critical power $P_s$ and the propagation constant
$b$. The solid line is the theoretical results and the square
symbols are the numerical results.} \label{fig2}
\end{figure}

Figure \ref{fig2}(a) shows the relation between the beam width and the critical power of
surface solitons. The solid line is the theoretical
value calculated by Eq. (\ref{5}) and the square symbols are the
numerical value. It can be seen that the theoretical value is in
good agreement with the numerical value. The relation between the
beam width and the propagation constant $b$ is shown in Fig.
\ref{fig2}(b). One can see that there exists a great difference between
the analytical solution and the numerical results. It means that Eq.
(\ref{447}) is incorrect because we neglect the higher order terms
of the expansion of the nonlinear refractive index in Eq.
(\ref{444}). The higher order terms  have little influence on the beam
intensity and the critical power, but a great deal of influence on the
phase change and the propagation constant. Because of the same
reason, the dependence of the critical power $P_s$ on the propagation
constant $b$ does not agree with the numerical solution, as shown in
Fig. \ref{fig2}(c).

\begin{figure}[htbp]
  \includegraphics[width=7.6cm]{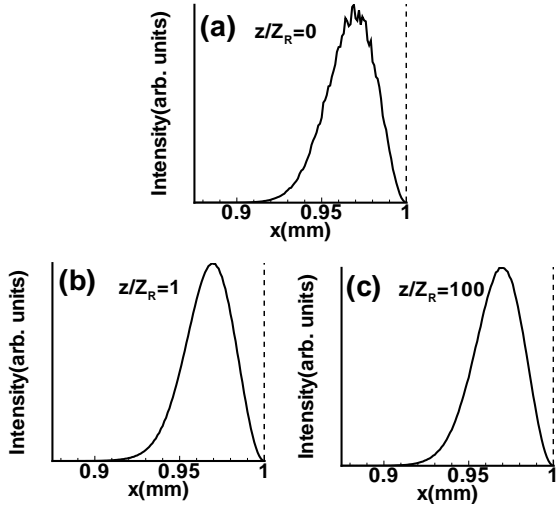}\\
  \caption{(a) The initial intensity distribution with $5\%$ noise.
 The intensity distributions of (b) after propagating $1$ Rayleigh distance and
(c) after propagating $100$ Rayleigh distances. The beam width is
$30$ $\mu$m. } \label{fig33}
\end{figure}

In order to confirm the stability of our solutions, we simulate the
beam propagation with perturbation based on Eqs.
(\ref{200})-(\ref{01}). The input condition is $U_t=U+\Delta
U=U[1+\epsilon R(x)]$, where $\Delta U$ is the perturbation
amplitude, $U_t$ is the total amplitude, $U$ is given by Eq.
(\ref{3}). $\epsilon$ is a perturbation constant, which is $5\%$ in our
simulation. $R(x)$ is a random function whose value is between $0$
and $1$. Figure \ref{fig33}(a) shows the initial intensity
distribution of the surface soliton with a great distortion. After
propagating about $1$ Rayleigh distance as shown in Fig.
\ref{fig33}(b), the distortion disappears and the intensity profile
becomes smooth. Figure \ref{fig33}(c) shows the intensity distribution
after propagating $100$ Rayleigh distances. This intensity
distribution is still smooth and without distortion. Compared with
Fig. \ref{fig33}(b), there is little change of the optical intensity
profile in Fig. \ref{fig33}(c). So, the distorted profile can reform
itself into a proper soliton after a short distance ($1$ Rayleigh
distance) and propagate along the sample surface stably.

\begin{figure}[htbp]
  \includegraphics[width=7.6cm]{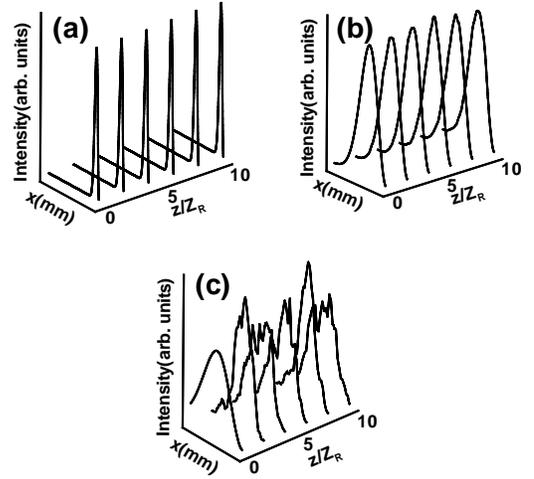}\\
  \caption{The sectional profile of the optical intensity of the surface soliton propagation
  for different sample widths. (a) The sample width $d=0.5$ mm $(\sim16w_0)$ , (b) $d=0.09$ mm $(3w_0)$, and
 (c) $d=0.06$ mm $(2w_0)$. The beam width is $30$ $\mu$m.}\label{fig5}
\end{figure}

Finally, we discuss the influence of the sample width on the surface
solitons. Fig. \ref{fig1}(b) shows the propagation of the surface
soliton for the beam width $w_0=30$ $\mu$m and the sample width $d=1$ mm
$(\sim33w_0)$, which is far greater than the beam width. If
the sample width is reduced to $d=0.5$ mm $(\sim16w_0)$  as shown in Fig.
\ref{fig5}(a), the propagation of the surface soliton is not
influenced by the left boundary. And if the sample width  is reduced to
$d=0.09$ mm which is three times of the soliton beam width, the
influence of the left boundary is still small. However when the sample width
is reduced to $d=0.06$ mm $(2w_0)$ as shown in Fig. \ref{fig5}(c), the
propagation of the surface soliton is unstable and there is a great
influence of the left boundary on the optical beam, and the optical beam
can not form a stable soliton. We take the three times of beam width
as a critical value. When $d/w_0 = 3$, the optical intensity at left
boundary is about $I|_{x=0}/I_{max}\approx3\times10^{-3}$, and the influence of the left boundary is small enough to be negligible.
When the sample width is smaller than three times of the beam width,
$I|_{x=0}$ is not negligible, e.g. $I|_{x=0}/I_{max}\approx0.1$ for
$d/w_0 = 2$, and surface soliton can not exist due to the influence
of the left boundary. This property is identical with that of
solitons in bulk media \cite{Dong-2010-PRA}.

\section{surface breathers}

\begin{figure}[htbp]
 \includegraphics[width=7.6cm]{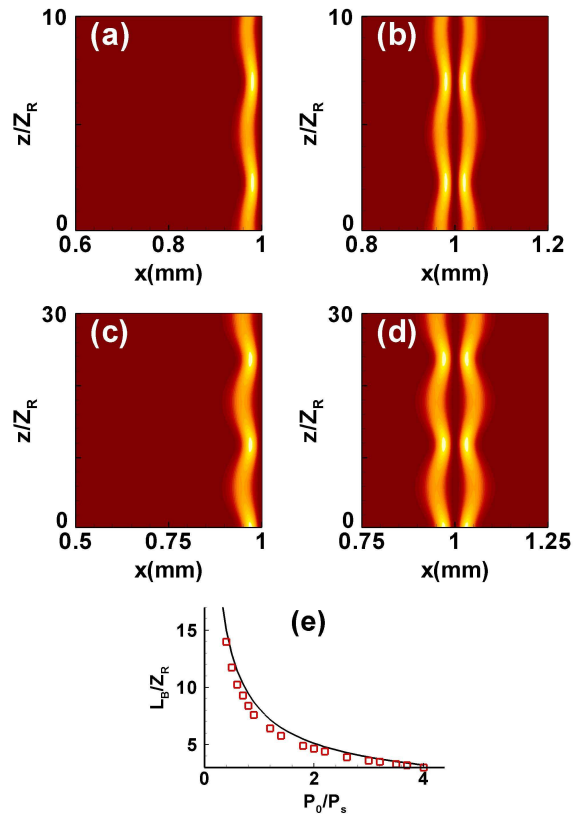}\\
 \caption{(Color online) The propagations of surface breathers and bulk
 breathers. (a)and (c), surface breathers with $P_0=2P_s$ and $P_0=0.5P_s$, respectively. (b)and (d), HG breathers with  $P_0=2P_c$ and $P_0=0.5P_c$, respectively.
 The initial beam width is $w_0=30\mu m$ for all. (e) The relation between the surface breathing
 period and $P_0/P_s$ for the analytical value (solid line) and the numerical
 value (square symbols).} \label{fig4}
\end{figure}

An optical beam can form a breather when the input beam power
is stronger or weaker than the soliton critical power in the bulk medium.
Fig. \ref{fig4}(b) ($P_0=2P_c$) and Fig. \ref{fig4}(d) ($P_0=P_c/2$)
show the first-order HG breathers with $w_0=30$ $\mu$m in bulk
media. Since the antisymmetry of the first-order HG beams in bulk
media can retain during propagation, it is reasonable to expect that
an optical beam launched at the sample surface also owns breathing
effect. When the surface beam power is stronger than the soliton
critical power, i.e. $P_0=2P_s$ as shown in Fig. \ref{fig4}(a), the
optical beam approaches to the interface at first and converges because
of the limit of the boundary. As the beam arrives at a
position near the interface, it is reflected by the interface and
diverges again. When the beam is far away from the interface, it is
attracted again to the interface. This property is identical with
the first-order HG breather in the bulk medium shown in Fig.
\ref{fig4}(b) at $0\leq x\leq d$. If the surface beam power is
weaker than the soliton power, i.e. $P_0=P_s/2$ as shown in Fig.
\ref{fig4}(c), the breathing effect is opposite to the case of the
stronger power. The optical beam is reflected by the interface and
diverges at first, then attracted by the interface and returns back.

The analytical solution of surface breathers can be also obtained
from the result of bulk breathers. In the bulk medium, we search for a
breather solution for Eq. (\ref{1}) in the HG function form
\begin{equation}\label{400}
U(x)=U_0H_1\left[\frac{(x-d)}{w(z)}\right]\exp\left[-\frac{(x-d)^2}{2w(z)^2}+ic(z)(x-d)^2\right]e^{-ibz},
\end{equation}
where $c(z)$ is the wavefront curvature. Substituting the breather
solution into Eq. (\ref{446}), and from the
coefficient of each order term of $x$, we obtain two equations \cite{Guo-2004-PRE}
\begin{subequations}
\begin{equation}\label{412}
\frac{dw(z)}{dz}-\frac{2c(z)w(z)}{k_1^2}=0,
\end{equation}
\begin{equation}\label{413}
\frac{dc(z)}{dz}-\frac{1}{2k_1w(z)^4}+\frac{2c(z)^2}{k_1}+\frac{k_1}{2}\left(\frac{\alpha\beta
}{n_1\kappa}\right)\frac{P_0}{2\sqrt{\pi}w(z)}=0.
\end{equation}
\end{subequations}
The combination of Eq. (\ref{413}) with the derivative form of Eq.
(\ref{412}) yields
\begin{equation}\label{4133}
\frac{d^2[w(z)/w_0]}{d(z/Z_R)^2}-\frac{1}{2[w(z)/w_0]^3}+\frac{P_0}{2P_c}=0,
\end{equation}
Using the method introduced in Ref. \cite{Guo-2004-PRE}, we obtain
\begin{equation}\label{414}
w(z)=w_0\left[\sqrt{\Lambda}+\sqrt{\frac{5\Lambda(\Lambda-1)}{3}}
\cos\left(\sqrt{\frac{3}{2}}\frac{z}{\Lambda Z_R}\right)\right],
\end{equation}
where $\Lambda=(2P_c/P_0)^{2/3}$. Then the parameters $c(z)$ and $b$ can
be gotten and the solution of HG breathers in bulk media is obtained.
The analytical solution of surface breathers is just the half
part of Eq. (\ref{400}), i.e. $0\leq x\leq d$. It should be noted that the critical power of surface breathers is a half of that of bulk solitons, i.e. $P_s=P_c/2$ .

From Eq. (\ref{414}), the breathing period $L_{B}$ can be obtained as
\begin{equation}\label{415}
L_{B}=\frac{2\sqrt{6}\pi Z_R}{3}\left(\frac{2P_c}{P_0}\right)^{2/3}.
\end{equation}
A surface breather has a same period as its bulk
counterpart. In Fig. \ref{fig4}(e), the analytical relation (solid
line) approximates well to the numerical results (square symbols).

\section{surface multipole solitons}

\begin{figure}[htbp]
  \includegraphics[width=7.6cm]{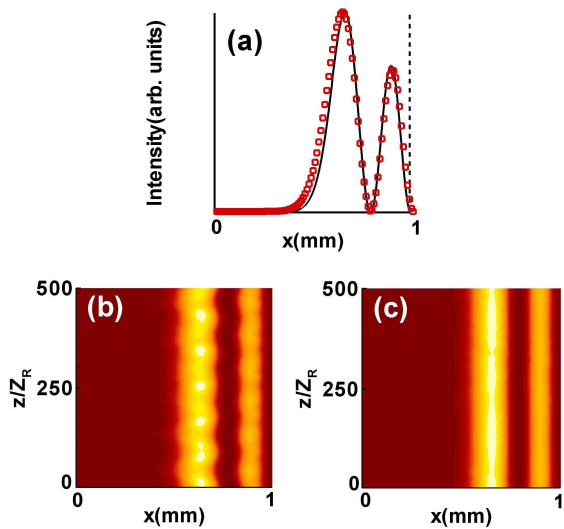}\\
  \caption{(Color online) (a) The profile of a dipole surface soliton. The
  square symbols are the numerical solution and the solid line is the
  analytical solution. (b) Propagation of the dipole soliton using the analytical solution as the incident
  profile. (c)Propagation of the dipole soliton using the numerical solution as the incident
  profile.
  The sample width is $1$ mm and the beam width is 250 $\mu$m. }\label{second}
\end{figure}

In this section, we discuss the surface multipole solitons using the
same method mentioned above. A surface dipole soliton can be
expressed as a half part ($0\leq x\leq d$) of a third-order HG
soliton in the center of a bulk medium ($0\leq x\leq 2d$). The
profile of a dipole soliton (solid line) is shown in Fig. \ref{second}(a), and
the numerical iterative solution (square symbols)  is also shown for comparing. It can
be seen that there exists a little difference between the numerical
solution and the analytical solution.
It is known that surface dipole solitons are stable \cite{V.Kartashov-2009-OL}, analogously
quadropole solitons in bulk media are stable too \cite{Dong-2010-PRA} .
The difference between the numerical solution and the
approximate analytical solution can be considered as a
perturbation, which leads to some spatial oscillations  in  propagation when the analytical solution is used as the initial profile, as shown in Fig.
\ref{second}(b). However, the propagation using the numerical
solution is stable as shown in Fig.
\ref{second}(c).

\begin{figure}[htbp]
  \includegraphics[width=7.6cm]{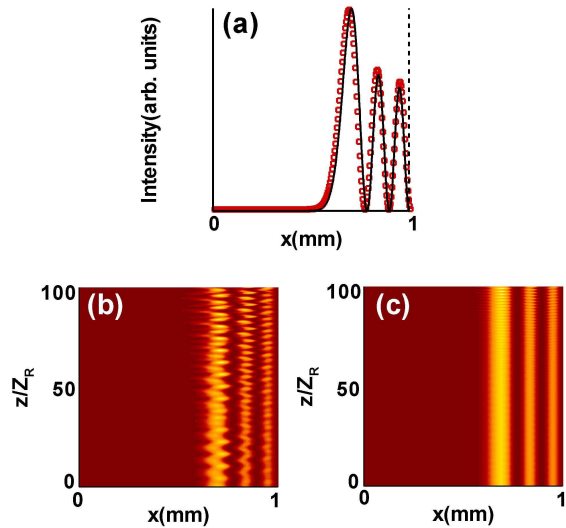}\\
  \caption{(Color online) (a) The profile of a tripole surface soliton. The
  square symbols are the numerical solution and the solid line is the
  analytical solution. (b) Propagation of the tripole soliton using the analytical solution as the incident profile. (c)Propagation of the tripole soliton using the numerical solution as the incident
  profile.
  The sample width is $1$ mm and the beam width is 120 $\mu$m.}\label{3jie}
\end{figure}

A surface tripole soliton can be expressed as a half part ($0\leq
x\leq d$) of a fifth-order HG soliton in the center of a bulk medium
($0\leq x\leq 2d$). The profile of a tripole soliton (solid line) is shown in
Fig. \ref{3jie}(a), and the numerical iterative solution (square
symbols) is also shown for comparing. There also
exists a little difference between the numerical solution  and the analytical solution. It has been proved in Ref.
\cite{V.Kartashov-2009-OL} that surface tripole solitons are unstable. As shown in  Fig.\ref{3jie}(c),
the oscillation occurs after  propagating over relatively long distance, i.e. $z/Z_R \approx 100$, when we use the numerical solution as the incident profile. When we use the approximate analytical solution,  the perturbation resulted from the difference between the numerical solution and the
approximate analytical solution grows up explosively. As a result, the strong oscillation takes place during the propagation as shown in Fig. \ref{3jie}(b). Anyway, our analytical solution coincides approximately with the numerical solution and it will be helpful
for analyzing the multipole surface soliton propagation.

\section{conclusion}
In conclusion, the analytical solution for surface fundamental
solitons has been found by comparing with bulk solitons, also
for surface breathers, surface dipole and tripole solitons. The
key assumption in this paper is that the optical fields all locate
in the nonlinear medium when two media have large difference in the
linear refractive index, and the intensity at the interface can be
negligible as $I=0$. This method is similar to the image beam method
\cite{Shou-2009-OL} in dealing with the boundary force exerted on
spatial solitons, which is an analogue of the well-known image
charge method. By introducing image beams, the boundary conditions are
automatically satisfied and analytical solutions can be found
conveniently. The numerical simulations show our assumption is
reasonable and correct. This method can be used also in dealing with other
surface solitons if the linear refractive index difference between
two media is large enough.

\section*{Acknowledgements}
This research was supported by the National Natural Science
Foundation of China (Grant Nos.10804033 and 10674050), the Program
for Innovative Research Team of Higher Education in Guangdong (Grant
No.06CXTD005), and the Specialized Research Fund for the Doctoral
Program of Higher Education (Grant No.200805740002).


\begin{thebibliography}{99}

\bibitem{Snyder-1997-Science}
A. W. Snyder and D. J. Mitchell, Science {\bf276}, 1538 (1997).

\bibitem{Krolikowski-2000-PRE}
W. Kr\'{o}likowski and O. Bang, Phys. Rev. E {\bf63}, 016610 (2000).

\bibitem{Conti-2004-PRL}
C. Conti, M. Peccianti, and G. Assanto, Phys. Rev. Lett. {\bf92},
113902 (2004).


\bibitem{Rotschild-2005-PRL}
C. Rotschild, O. Cohen, O. Manela, M. Segev, and T. Carmon, Phys.
Rev. Lett. {\bf95}, 213904 (2005).

\bibitem{Deng-2007-JOSAB}
D. Deng, X. Zhao, Q. Guo, and S. Lan, J. Opt. Soc. Am. B {\bf24},
2537 (2007).

\bibitem{Xu-2005-OL}
Z. Xu, Y. V. Kartashov, and L. Torner, Opt. Lett. {\bf30}, 3171
(2005).

\bibitem{Rotschild-2006-OL}
C. Rotschild, M. Segev, Z. Xu, Y. V. Kartashov, L. Torner, O. Cohen,
Opt. Lett. {\bf31}, 3312 (2006).

\bibitem{Buccoliero-2007-PRL}
D. Buccoliero, A. S. Desyatnikov, W. Krolikowski, and Y. S. Kivshar,
Phys. Rev. Lett. {\bf98}, 053901 (2007).

\bibitem{Alexander-2005-PRE}
A. I. Yakimenko, Y. A. Zaliznyak, and Y. Kivshar, Phys. Rev. E
{\bf71} 065603(R) (2005).

\bibitem{Briedis-oe-2005}D. Briedis, D.E. Petersen, D. Edmundson, W. Krolikowski and O. Bang,
 Opt. Express, {\bf 13},435-443(2005).

\bibitem{Tomlinson-ol-1980}W. J. Tomlinson,
 Opt. Lett. {\bf 5}, 323-325 (1980).

\bibitem{Mihalachea-pio-1989}D. Mihalachea, M. Bertolottib and C. Sibiliab,
Prog. Opt.  {\bf 27}, 227-313(1989).

\bibitem{Makris-ol-2005}K. G. Makris, S. Suntsov, D. N. Christodoulides, G. I. Stegeman, A. Hache,
 Opt. Lett. {\bf 30}, 2466-2468 (2005).

\bibitem{Suntsov-prl-2006}S. Suntsov, K. G. Makris, D. N. Christodoulides, G. I. Stegeman, A. Hache, R. Morandotti, H. Yang, G. Salamo, and M. Sorel, Phys. Rev. Lett. {\bf 96}, 063901 (2006).

\bibitem{Katashov-prl-2006}Y. V. Kartashov, V. A. Vysloukh, and L. Torner,
 Phys. Rev. Lett. {\bf 96}, 073901 (2006).

\bibitem{Quirino-pra-1995}G. S. Garcia Quirino, J. J. Sanchez-Mondragon, and S. Stepanov ,
Phys. Rev. A {\bf 51}, 1571-1577 (1995).

\bibitem{Cronin-Golomb-ol-1995}M. Cronin-Golomb, Opt. Lett. {\bf 20}, 2075-2077 (1995).

\bibitem{Aleshkevich-pre-2001}V. Aleshkevich, Y. Kartashov, A. Egorov,  and V. Vysloukh,
Phys. Rev. E {\bf 64}, 056610 (2001).

\bibitem{Lazarides-pre-2008}N. Lazarides, G. P. Tsironis, and Y. S. Kivshar,
Phys. Rev. E {\bf 77}, 065601(R)(2008).

\bibitem{Alfassi-2007-PRL}
B. Alfassi, C. Rotschild, O. Manela, M. Segev, and D. N.
Christodoulides, Phys. Rev. Lett. {\bf98}, 213901 (2007).


\bibitem{Alfassi-2009-PRA}
B. Alfassi, C. Rotschild, and M. Segev, Phys. Rev. A {\bf80}, 041808
(2009).

\bibitem{Ye-2008-PRA}
F. Ye, Y. V. Kartashov, and L. Torner, Phys. Rev. A {\bf77}, 033829
(2008).

\bibitem{V.Kartashov-2009-OL}
Y. V. Kartashov, V. A. Vysloukh, and L. Torner, Opt. Lett. {\bf34},
283 (2009).



\bibitem{Huang-oc-2006}Y. Huang, Q. Guo, and J. Cao,
Opt. Commun. {\bf 261}, 175-180(2006).

\bibitem{Cao-oc-2008}L. Cao, Y. Zhu, D. Lu, W. Hu, and Q. Guo,
 Opt. Commun. {\bf 281}, 5004-5008(2008).



\bibitem{Dong-2010-PRA}
L. Dong, and F. Ye, Phys. Rev. A {\bf81}, 013815 (2010).

\bibitem{Guo-2004-PRE}
Q. Guo, B. Luo, F. Yi, S. Chi, and Y. Xie, Phys. Rev. E {\bf69},
016602 (2004).



\bibitem{Shou-2009-OL}
Q. Shou, Y. Liang, Q. Jiang, Y. Zheng, S. Lan, W. Hu and Q. Guo,
Opt. Lett. {\bf34}, 3523 (2009).


\end{thebibliography}
\end{document}